# 2025 Quantum Diamond Workshop

Findings Report

Danielle A. Braje[1,*], Matthew L. Markham[2,†], Jennifer M. Schloss[1,‡], Michael A. Slocum[3,§], Ronald L. Walsworth[4,‖]

## Overview and Goals

This report synthesizes the outcomes of a two-day workshop held in Washington, D.C. in May, 2025 that convened researchers, industry representatives, and government stakeholders to examine the current state and future directions of quantum diamond technologies. The workshop's goals were to assess the most promising use cases, to identify the key technical and structural challenges limiting adoption, and to chart potential pathways for aligning application needs with diamond material and device development. Through a series of technical presentations and open discussions, participants explored both near-term demonstrations and long-term infrastructure needs, highlighting the critical role of coordination between material suppliers, device engineers, and end users. The goal of this report is to distill those insights into a coherent set of cross-cutting themes, challenges, and strategic actions that can guide government, industry, and academic efforts to accelerate the maturation and commercialization of quantum diamond technologies.

## Executive Summary

The field of quantum diamond technologies has reached a pivotal juncture (Barry et al., 2020; Degen et al., 2017). Across sensing, microscopy, networking, and materials science, academic groups and government programs have demonstrated impressive capabilities—flight tests of vector magnetometers for magnetic navigation, NV microscopes that outperform SQUIDs, quantum memories for scalable computing architectures, and initial single-molecule detection. Many of these have achieved mid-level technology readiness (TRL 5 or higher). These accomplishments underscore the breadth of potential impact, from defense navigation and microelectronics diagnostics to geoscience and biomedical sensing (Aslam et al., 2023). Yet despite the technical progress, commercial adoption remains limited: application pull is uneven, markets are fragmented, and investment across industry lacks alignment and has gaps at key points, including at the seed stage.

The overarching challenges are as much structural as they are scientific. In many application areas, quantum diamond material is not the bottleneck—adoption depends instead on integration into established workflows, regulatory acceptance, user familiarity, or investment to

1. MIT Lincoln Laboratory, 2. Element Six, 3. Air Force Research Laboratory, 4. University of Maryland

[*]braje@ll.mit.edu, [†]matthew.markham@e6.com, [‡]jennifer.schloss@ll.mit.edu, [§]michael.slocum.2@us.af.mil, [‖]walsworth@umd.edu

scale established solutions. Where diamond materials are decisive, the supply chain remains fragile: high-quality seeds, isotopically pure methane, irradiation and annealing services, and high-throughput characterization are limited, fragmented, or costly. Standardization is both necessary and elusive. Without clear mappings from application requirements to objective material properties, developers struggle to specify what they need and suppliers struggle to scale. The community risks remaining in a chicken-and-egg cycle: applications cannot advance without defined material products, and materials cannot mature without application pull.

Strategic progress will require both ambition and pragmatism. At the ambitious end, large-scale government and industry investment can accelerate development of exemplar systems—diamond magnetometers with deployable SWaP, microscopy tools integrated into semiconductor fabs, and quantum interconnect nodes with engineered emitters—that not only deliver immediate utility but also generalize infrastructure for other applications. At the pragmatic end, the community can take action now by agreeing on provisional benchmarks, iterating standards as experience grows, and convening structured dialogue among material suppliers, device engineers, and end users. Spiral development models—advancing device design, fabrication processes, and the material platform in parallel—offer a practical way to shorten cycles and reduce cost. Dedicated programs to strengthen the domestic supply chain, coupled with patient capital for startups and targeted SBIR-style support, can sustain early commercialization. Taken together, these actions will bridge the gap between promising academic demonstrations and sustainable commercial adoption, ensuring that diamond technologies move from boutique successes to broadly impactful platforms for defense, industry, and society.

**Key Takeaways and Strategic Actions**

- Quantum diamond technologies have relatively high maturity (TRL 5 or higher), with a strong base of past academic research, promising applications in diverse sectors, and early-stage commercialization underway, especially in microscopy and networking.
- However, large-scale commercial adoption is lagging, largely due to lack of a clear demand signal and relatively low investment in recent years from both government and the private sector.
- As a result, progress with maturing quantum diamond material and the supporting technology supply chain remains stuck in a chicken-and-egg cycle.
- Focused government and industry investment is needed to break out of this cycle and accelerate development of exemplar systems that both provide near-term utility and enable infrastructure for other applications.
- Continued community efforts are also needed to establish technology benchmarks and standards, and encourage dialogue among material suppliers, device engineers, and end users.

# Use Cases & Needs

To date significant advancements have been made in the development of sensors and quantum computing subsystems using diamond (Doherty et al., 2013; Barry et al., 2020). At a high level, this technology has applicability in a wide number of market sectors such as defense, aerospace, healthcare, materials, energy, communications and computing (Schirhagl et al., 2014; Awschalom et al., 2021; Kitching et al., 2023; Budakian et al., 2024). However, much of the early adoption of diamond-based solutions has been developed through a technology push framework, since the solutions are typically for new applications rather than replacing existing solutions with improved capabilities. The result has been multiple demonstrations that have reached a technology readiness level (TRL) of 5 or higher, which include flight demonstrations of nitrogen-vacancy (NV) magnetometers for magnetic navigation (MagNav), quantum memories that have the potential for integration into larger quantum computing systems, NV microscopes that can image magnetic fields from integrated circuits (ICs) or geological specimens, and single-molecule localization and detection. Although these demonstrations have been impressive, there are still technological and commercial gaps to overcome before there is widespread adoption.

The Micro-Electronic Commons technology area leads for quantum technology explicitly called out a 'diamond foundry platform' and 'quantum memories' as technology gaps, which would unlock scalable fab transitions. It is expected that the maturation of these competencies and fabrication capabilities will continue to expand the availability of materials and subsystems that can be integrated into future quantum technologies. Discussion for this session focused on the need to identify specifications for use cases. A challenge to providing these specifications is that diamond technologies typically do not serve as drop-in replacements for existing technologies but rather leverage the unique behavior of quantum sensors to provide new and different capabilities.

It is expected that early adopters will often be defense related due to the appetite to pay more for exquisite technology, and the ability to accept risk. Initial applications with potential to demonstrate utility include MagNav, integrated circuit (IC) inspection tools, or IC inspection that enables verification for Trusted & Assured Microelectronics, as well as components for quantum computers, such as memory. In each case there is some level of the need to co-develop the diamond-based technology along with the application to realize a new capability. To achieve this, it is essential that technologists working with diamond develop tight relationships with teams working on application to ensure that an understanding of the application requirements drives diamond technology development. The energy sector likely has significant demand for advanced sensing relevant to diamond since there is a high need for low-drift E/M field sensors. If successful, the energy sector can drive significant scale to help reduce the cost for material and sensors due to increased production volume. This scaling will in part be made possible through developing integration techniques for the solid-state sensor material, which motivates investment in diamond integrated photonics. It is expected that integrated photonic devices will not only reduce SWaP, but also reduce cost, increase production volumes and facilitate adoption into more communities.

# Magnetometry

Diamond-based magnetometers show promise to impact a wide range of use cases from navigation and microelectronics to medicine and emerging technologies (Barry et al., 2020). Diamond magnetometers have demonstrated multiple unique strengths—including inherent vector output, intrinsic stability tied to physical constants, and inherently dead-zone-free operation. Additional properties expected in future prototypes include high sensitivity, broad dynamic range, and chip-scale integration, which could open new applications such as anomaly detection and tensor gradiometry, not easily matched by competing platforms. However, displacing existing technologies requires not just a promise of enhanced capabilities but also a combination of increased device performance or functionality with a path toward manufacturable technology. To this end, quantum diamond magnetometers are already demonstrating tangible advantages over existing technologies, with recent field trials showing that diamond vector magnetometers can more than double MagNav accuracy compared to scalar devices. Moreover, recent advances show potential for chip-scale integration of diamond devices, which would drive down SWaP and cost of deployed diamond magnetometers. These developments suggest that diamond magnetometry is moving beyond laboratory prototypes toward application impact.

Despite the progress, demonstrations of sensors with excellent sensitivity and stability have not yet translated into commercial products because the pathways to scale are constrained by both materials and system integration challenges. To accelerate development of commercial diamond sensors, diamond material growth and processing improvements are essential: yield improvements in etching, polishing, and surface processing are urgently needed, and better control of NV density, orientation, and subsurface damage would reduce variability in performance (Achard et al., 2019). However, due to a diversity of diamond requirements for different sensor modalities, no consensus has emerged on standard categories of "sensor-grade" diamond, making it difficult for suppliers to scale production efficiently. Furthermore, realizing the potential of ubiquitous chip-scale diamond magnetometry requires development of advanced semiconductor fabrication processes and techniques to enable the integration and manufacturability needed to realize the expected substantial SWaP-C reduction. It is expected that semiconductor manufacturing techniques developed for the novel use case of quantum interconnects could significantly accelerate the development of chip-scale diamond magnetometers, making coordination between teams and research efforts paramount (Agio & Castelletto, 2025).

Addressing these issues requires a coordinated development pathway. Standardizing material specifications would allow sensor developers to design against known targets while enabling suppliers to scale. Achieving this end will require increased coordination between sensor designers and material manufacturers to clarify the mapping from diamond properties to magnetometer design requirements that could simplify diamond manufacturing and streamline investment. Spiral development strategies, where sensor architectures, fabrication flows, and material improvements are advanced in parallel, offer a pragmatic way to shorten iteration cycles and reduce cost. National investment in quantum diamond material production, processing, and benchmarking capacity would accelerate sensor development, provide confidence in device performance, and de-risk industrial adoption. Taken together, these steps could move magnetometry from promising prototypes toward deployable, low-SWaP instruments suitable for both defense and commercial markets.

# Microscopy

Diamond-based magnetic microscopy is arguably the most commercially advanced of the NV sensing applications, with both nano- and micron-scale resolution instruments already in use across scientific domains (Levine et al., 2019). In geoscience, for example, NV microscopes already outperform SQUIDs in spatial resolution, and with improved sensitivity could become the first all-purpose magnetic probe for natural samples. In microelectronics, NV microscopy provides a unique combination of capabilities not available with traditional techniques—including the ability to image currents through opaque layers at room temperature over a wide field of view—which makes it attractive for semiconductor process control, failure analysis, and supply chain assurance. Discussions underscored how magnetic current imaging has revealed faults in advanced packages that elude other diagnostic tools, highlighting how diamond's unique properties map onto urgent needs in heterogeneous integration and advanced packaging.

The barriers to broader adoption are more practical than conceptual. Instruments today remain largely bespoke, built from custom components by expert groups, which limits scalability and increases cost. For microelectronics and biomedical markets, which are highly conservative in adopting new tools, convincing evidence of utility must be demonstrated—with data integrated into existing workflows—before significant uptake occurs. Meeting this high bar requires application-tailored development. Such development must improve not only microscope performance but also demonstrate how new capabilities—through-package measurements, vector magnetic imaging, or real-time current mapping—should be interpreted and acted upon. This is a dual challenge: instrumentation development must advance hand-in-hand with application co-development so that end users understand the impact of new modalities that have no direct equivalents in traditional metrology. In the U.S., the commercialization pipeline is thin, with few startups pursuing diamond microscopy compared to the vibrancy seen in quantum computing and networking. Federal support for academic research has also declined relative to a decade ago, while Europe and Asia are investing more steadily in both science and commercialization. Despite pioneering much of the early work and intellectual property, the U.S. risks slipping behind in both academic leadership and market presence.

Strategic progress will require addressing both technical and structural issues. On the technical side, performance targets need to be explicitly tied to use cases, with application-specific benchmarks for per-pixel sensitivity, field-of-view, dynamic range, and compatibility with biological, geological, or electronic samples. These should be framed not just around what diamond microscopes can do today, but also around what new modalities mean for end-user workflows. On the structural side, both ideal and more pragmatic pathways forward can be articulated. Ideally, large-scale government investments could accelerate progress by funding major programs to establish application-specific benchmarks—such as per-pixel noise floors, field-of-view, dynamic range, and sample compatibility across biological, geological, and electronic systems—and by sustaining long-term academic and interdisciplinary research that matures both instruments and use cases. In parallel, pragmatic steps the community can take now include defining provisional benchmark metrics—even if imperfect—and iterating them as experience grows; convening end users and developers to better understand how new capabilities can be leveraged; and working toward standardized subsystems (NV layers, photonics, MW antennas, mounts) that reduce reliance on bespoke builds. Patient capital and SBIR-style programs could help U.S. startups survive the long development timelines, while pilot deployments in semiconductor fabs, supported, if necessary, by national security programs, could validate utility and de-risk adoption in conservative industries. Taken together, these actions would not only mature the microscopes themselves but also strengthen the understanding of their utility, ultimately expanding diamond microscopy from boutique scientific instruments into widely adopted platforms for microelectronics, defense, and beyond.

## Emerging Uses

Recent work has highlighted new frontiers where diamond may play an essential or supporting role, underscoring both its versatility and the need for strategic focus. In quantum interconnects, group-IV vacancy centers coupled to diamond photonics are among promising candidates for spin-photon interfaces, a uniquely enabling role where diamond faces little competition (Awschalom et al., 2021). Diamond development in this space focuses on diamond fabrication and heterogeneous integration. In extreme environments, such as high-pressure experiments probing superconductivity, diamond again offers unique advantages, though there are challenges in sourcing material with the required properties. In biomedicine, by contrast, bottlenecks are more often in system integration, biological compatibility, and regulatory approval than in the diamond material itself, and some promising work, such as protein-based qubits, has drawn inspiration from NV sensing while bypassing diamond altogether.

The common theme across these domains is that vetting new ideas takes time and requires sustained investment in multiple domains to translate techniques that show early promise to a product with practical benefit. Given the broad potential of diamond quantum technologies, there is a wide diversity of applications; however, this can result in splitting of resources. Moreover, while academic groups can pioneer novel modalities, they often cannot provide the engineering maturity, system integration, or market alignment required to carry concepts forward. As a result, many emerging uses risk remaining promising demonstrations rather than scalable technologies.

Photonic interconnect nodes with controlled emitter properties exist in a unique place within the diamond ecosystem, having strong scientific and application pull, leveraging the significant investor interest in quantum computing, utilizing commercially available diamonds, and driving development of critical infrastructure like diamond photonic integration. The willingness to take significant risk to mature interconnects demonstrates a model for aligned government and private investment to develop technologies that provide a unique advantage while also maturing manufacturing and fabrication techniques that can be translated into other markets.

Sustained basic research funding is essential to nurture these exploratory directions, but must be paired with platform-level investments that can generalize to multiple applications. The science-first approach is essential to developing novel techniques. Moreover, the industry needs advocates who can foresee how scientific innovation will lead to new capabilities that spark interest and investment. This combination of infrastructure development, long-horizon support, and acceptance of risk offers a pathway to bring emerging uses from promising demonstrations to impactful technologies.

## Diamond Growth, Fabrication, and Enabling Technologies

The foundation of all quantum-diamond technologies is the ability to grow, process, and characterize material with properties tuned to specific applications (Achard et al., 2019, Barry et al., 2020; Agio & Castelletto, 2025). Decades of research have laid a strong base, and progress has been made on multiple fronts: reducing dislocations in bulk crystals, improving surface flatness and smoothness, mitigating subsurface damage, and refining nitrogen incorporation for both ensemble and single-defect applications. More recently, smart-cut membrane approaches are beginning to open pathways to thin-film devices analogous to CMOS. These methods point toward reproducible transfer of thin diamond layers onto larger substrates, enabling more complex integration with photonics and electronics. Efforts are also underway to leverage advances in large-area diamond growth—through heteroepitaxial and mosaic methods—to enable wafer-scale quantum diamond in the future, although scaling without introducing strain and defects remains a persistent challenge. Increasingly, sophisticated characterization methods such as birefringence mapping, ODMR imaging, and in situ growth monitoring are being used to close the loop between seed properties, growth processes, and application-level performance metrics, indicating a shift toward more systematic, data-driven materials development.

Despite these advances, the development of quantum-enabled diamond devices continues to face challenges. The diversity of application requirements makes streamlined optimization difficult: different use cases demand different balances of NV density, coherence time, isotopic enrichment, surface quality, or geometry. As a result, device developers often cannot simply order "quantum-grade" diamond in the same way that chip designers can order standardized silicon wafers. The supply chain is fragmented — high-quality seeds, isotopically pure methane, irradiation, annealing, cutting, polishing, and other post-processing steps are often sourced separately, introducing inefficiencies, delays, and cost. While bulk diamond products are produced at scale in the U.S. for non-quantum applications, no facility is currently equipped to produce functionalized diamond at scale for quantum applications with the necessary application-specific characterization. Even seemingly straightforward metrics like ODMR contrast depend strongly on measurement conditions, which undermines their usefulness for specifying or procuring material and makes it difficult for developers to translate application requirements into material specifications.

The absence of a shared framework further limits progress. While academic initiatives and integrated efforts such as those at MIT Lincoln Laboratory have shown that coupling synthesis through device testing can speed up iteration cycles, these models are not easily scaled across the industry. The broader value chain remains underdeveloped, with little standardization and few mechanisms for mapping application needs onto material products. This gap makes it difficult to identify commonalities where a single material type could serve multiple use cases, and it leaves suppliers uncertain about what products to scale.

Accelerating development will require both infrastructure and a new approach to standardization. Expanding domestic capacity across the pipeline—from seeds to growth, post-growth processing, thin-film transfer, and high-throughput characterization—would reduce dependence on fragile or international supply chains. More importantly, standardization should not be understood as narrowing the community to just a few diamond "SKUs," but rather as developing a clearer mapping between application requirements and objective material properties. This means linking performance across a range of applications to standardized figures of merit that can be measured reproducibly across different setups (e.g., $T_2^*$, NV concentration, strain uniformity, flatness, miscut angle). Establishing a standardized "Diamond Process Development Kit (PDK)," modeled after the semiconductor industry, could provide this structure. Such a PDK would define a small catalog of qualified material types, each with documented properties, measurement protocols, and recommended processes, while leaving room for iteration as needs evolve. It would also make explicit the feedback loop between application performance and material development, shortening cycles by ensuring that characterization feeds directly into growth recipes and device design.

Finally, the community will need to continue exploring ways to shorten development cycles that now span growth, processing, characterization, application testing, and feedback. While much of this may not be directly actionable in the near term, one concrete step is to foster regular, structured conversations among diamond engineers, device developers, and end-users to compare metrics, identify commonalities, and update specifications iteratively. Machine learning and data-driven approaches, such as those already being piloted, can help optimize growth and processing parameters more efficiently if they are paired with standardized inputs and outputs. Taken together, these actions would not only make diamond development more predictable and responsive to user needs, they would also give suppliers and investors greater confidence that the technology can scale beyond laboratory demonstrations.

# Acknowledgements

All authors contributed equally to this report. The authors are grateful to all attendees of the 2025 Quantum Diamond Workshop for their participation and thoughtful discussion, which formed the basis for this report. The authors also thank Matthew Turner for helpful comments on an earlier draft. This material is based upon work supported by the Department of the Air Force under Air Force Contract No. FA8702-15-D-0001 or FA8702-25-D-B002. Any opinions, findings, conclusions or recommendations expressed in this material are those of the author(s) and do not necessarily reflect the views of the Department of the Air Force. © 2025 Massachusetts Institute of Technology. Delivered to the U.S. Government with Unlimited Rights, as defined in DFARS Part 252.227-7013 or 7014 (Feb 2014). Notwithstanding any copyright notice, U.S. Government rights in this work are defined by DFARS 252.227-7013 or DFARS 252.227-7014 as detailed above. Use of this work other than as specifically authorized by the U.S. Government may violate any copyrights that exist in this work.